\documentclass[runningheads]{llncs}
\usepackage{graphicx}
\usepackage{makecell}
\usepackage{url}
\usepackage{xcolor}
\usepackage{color}
\usepackage[normalem]{ulem}
\begin{document}
\title{A Survey on Recent Advancements for AI Enabled Radiomics in Neuro-Oncology}
\author{Syed Muhammad Anwar \inst{1,3} \and
Tooba Altaf \inst{2} \and
Khola Rafique \inst{3} \and
Harish RaviPrakash\inst{1}\and
Hassan Mohy-ud-Din \inst{4}\and
%Syed Muhammad Anwar\inst{1,2}\orcidID{1111-2222-3333-4444} \and
Ulas Bagci\inst{1}}
\authorrunning{S. Anwar et al.}
\institute{Center for Research in Computer Vision, University of Central Florida, Orlando, FL, 32816, USA \and
Department of Computer Science, University of Wah, Wah, Pakistan \and
Department of Software Engineering, University of Engineering and Technology, Taxila, 47050 Pakistan \and
Syed Babar Ali School of Science and Engineering, Lahore University of Management Sciences (LUMS), Lahore, Pakistan\\
\email{s.anwar@knights.ucf.edu}\\
}
\maketitle              % typeset the header of the contribution
\begin{abstract}
Artificial intelligence (AI) enabled radiomics has evolved immensely especially in the field of oncology. Radiomics provide assistance in diagnosis of cancer, planning of treatment strategy, and prediction of survival. Radiomics in neuro-oncology has progressed significantly in the recent past. %Deep learning, a subset of machine learning, uses data for training neural networks in challenging application areas. 
Deep learning has outperformed conventional machine learning methods in most image-based applications. %Neural networks are flexible and can accommodate large number of parameters.
Convolutional neural networks (CNNs) have seen some popularity in radiomics, since they do not require hand-crafted features and can automatically extract features during the learning process. %A CNN has concatenated structure comprised of convolutional, pooling and fully connected layers. 
In this regard, it is observed that CNN based radiomics could provide state-of-the-art results in neuro-oncology, similar to the recent success of such methods in a wide spectrum of medical image analysis applications. Herein we present a review of the most recent best practices and establish the future trends for AI enabled radiomics in neuro-oncology. 

\end{abstract}
\section{Introduction}

Brain and other central nervous system (CNS) tumors account for the second most common cancer affecting children, and the third most common cancer affecting adolescents and young adults ~\cite{url2},~\cite{url1}. There are approximately 700,000 people with primary brain or CNS tumors in the United States alone~\cite{url2}. Treatment is dependent on multiple factors including age, gender, tumor size and location, etc. The standard approach in most cases is to surgically remove the tumor via craniotomy \cite{url3}. However, some tumors cannot be surgically removed and the treatment then relies on radiation therapy. Rigorous planning is necessary to determine the exact tumor volume and a buffer region surrounding the tumor which has to be treated to prevent growth from left over malignant cells. The accurate planning of resection and radiation area is challenging owing to the difficulty in determining the exact tumor dimensions. For manual segmentation (delineation), the radiologists need to carefully analyze a large amount of radiology images. To ease the load on radiologists, computational methods to automatically extract quantitative features (aka radiomics) from radiological scans  have been proposed. 

Radiomics comprises of numerous significant disciplines, including radiology, computer vision, and machine learning. The objective is the recognition of quantitative imaging features with an anticipation of significant clinical results in prognosis and analysis of certain treatment strategies \cite{zhou2018radiomics}.  The information provided by radiology scans is processed with the help of quantitative image analysis (QIA) for identifying patterns in radiology scans in a way that human eye may not achieve. Different steps includes, acquire and store images, segment and identify region of interest (ROIs), extract features, build and validate model, and integration of these process into clinical decision support system. The resultant units of data from QIA may be called quantitative imaging bio-markers depending on their predictive powers. A huge amount of information is captured during clinical imaging but the underlying data, in most cases, have been reported in subjective and qualitative terms. Specifically, radiomics in neuro-oncology aims to revamp the brain tumor treatment paradigm by extracting quantitative features from brain scans (MRI). %\cite{gillies2015radiomics}. 
Data is mined via multiple machine learning algorithms and can potentially be used as imaging bio-markers to distinguish intra-tumoral dynamics during treatment \cite{gillies2015radiomics}. With the increase in number of reported cancer cases, analytic methods for imaging have revealed new understandings about initial treatment response, risk factors, and optimal treatment approaches \cite{kotrotsou2016radiomics} \cite{zhou2017identifying}. Image-based models are turning into a significant empowering innovation that allow investigation and approval of selected quantitative features. 

The recent advancements, particularly in Artificial Intelligence (AI), are impacting major technological and scientific fields. To keep up with these advancements, medical science is adapting new methodologies for improving diagnosis and treatment of various clinical conditions \cite{lambin2017radiomics}. In clinical setting, imaging has played a vital role for a long time by helping physicians in diagnostic and treatment related decisions making \cite{aerts2014decoding}. However, over a passage of time, medical imaging has evolved from just being a diagnostic tool and is now beginning to take a critical role in precision medicine for tasks such as screening, diagnosis, guided treatment, and assessing the disease recurrence likelihood \cite{giardino2017role}. The emerging field of radiomics in oncology has helped in developing a latent solution for tumor characterization by extracting a large number of features from medical images \cite{kumar2012radiomics} \cite{parmar2014robust}. Attributes that can be used in assessment of tissue appearance by radiologist are of great importance and can be used in the development of medical imaging analysis techniques. Some common examples of such attributes include texture, intensity, and morphology. \textit{Texture} can be defined as the spatial variation of pixel intensities within an image, and is known to be particularly sensitive for the assessment of pathology \cite{gilanie2018classification}. Visual assessment of texture is however, particularly subjective. Additionally, it is known that human observers possess limited sensitivity to textural patterns, whereas computational texture analysis techniques can be significantly more sensitive to such changes \cite{fetit2018radiomics}. For image classification, numerous computer vision algorithms depend on extracting native characteristics form images.
%, that is the reason that most of the literature focused on finding , characterizing , and improving those features that were extracted form images. 
These features are handcrafted with an eye for resolving explicit issues like obstructions and variations in scale and brightness. The design of handcrafted features often involves finding the right trade-off between accuracy and computational efficiency \cite{nanni2017handcrafted}. In contrast, deep learning (DL) methods have a huge potential to replace conventional machine learning methods for automatically extracting imaging features which are more efficient and give state-of-the-art performance in a large number of applications already. In the following, we present a review of methods relying on handcrafted features and those using DL and analyze the future direction for AI enabled radiomics in neuro-oncology.

\section{Radiomics using Handcrafted Features}

\begin{figure}[!t]
    \centering
    \includegraphics[width = 140mm]{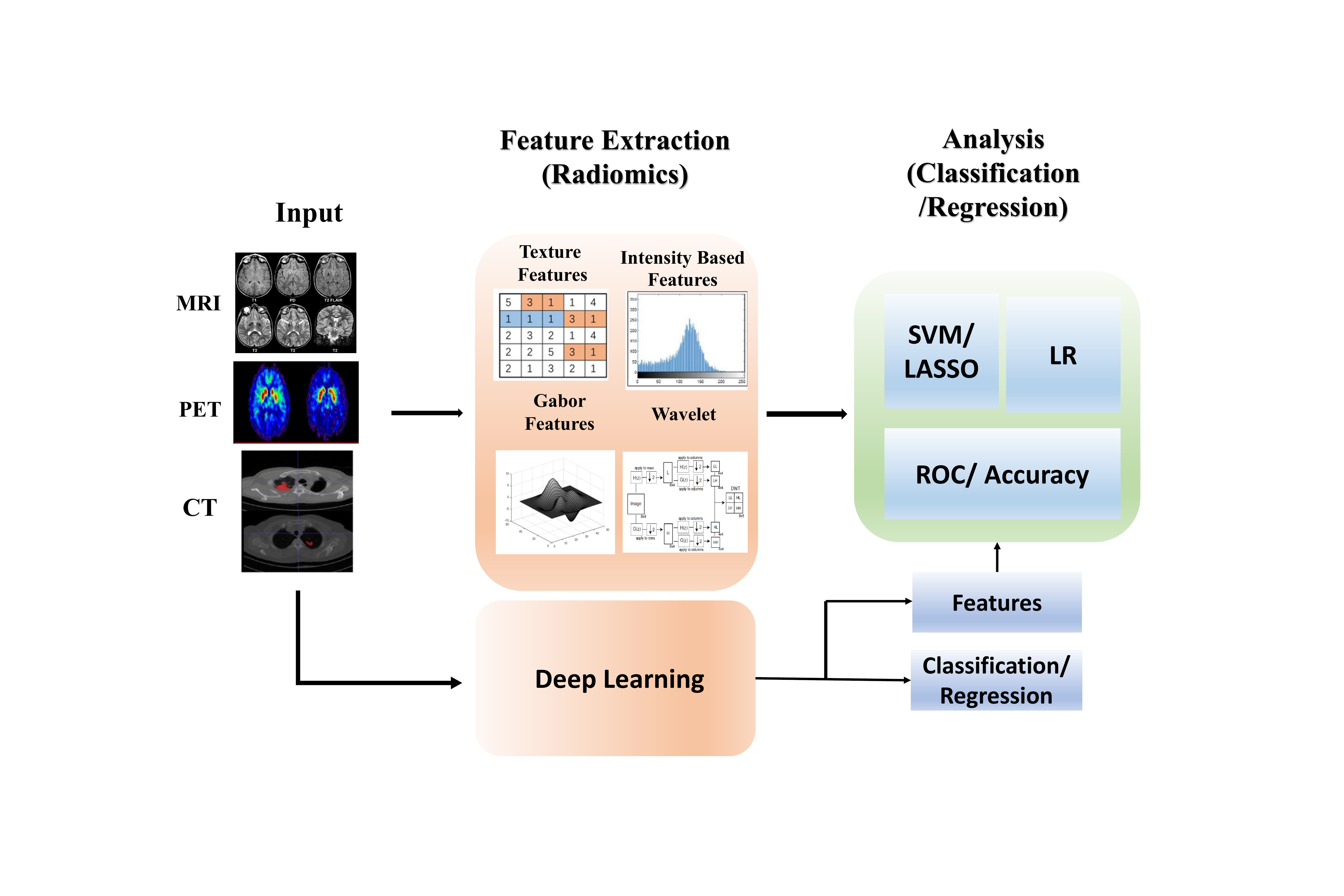}
    \vspace{-15mm}
    \caption{A pipeline of steps for radiomics in radiology using handcrafted and DL based features}
    \label{fig:my_label}
\end{figure}

%Table 1 show different state of the art techniques by using which 
A general pipeline for radiomics in neuro-oncology is shown in Figure \ref{fig:my_label}. Different radiomics features are extracted from medical images and then machine learning classifiers are used to detect diseases such as brain tumor. These radiomics features are either extracted in a hand crafted manner or through DL. The top layer (Figure \ref{fig:my_label}) shows how handcrafted features are used with different radiology image inputs. The feature extraction stage (also known as conventional radiomics approach) relies on selecting features from various domains such as texture, intensity/density, and frequency (e.g., wavelet). Different machine learning classifiers ( support vector machines (SVMs) and logistic regression(LR)) are used for analysis of these features and results are analyzed using performance parameters (such as accuracy and receiver operating characteristics (ROC)). Whereas, in deep learning the model chooses the appropriate features, allowing feature learning, which can then be used directly for classification/regression. These learned features can also be used with other classifiers such as SVM. %, lung cancer, bladder cancer and gastric cancer). %are some of the common examples shown in table.

One of the approaches employed to extract radiomic features is called local binary patterns (LBPs) \cite{abbasi2017detection} \cite{giacalone2018local} \cite{gupta2019glioma} \cite{polepaka2019idss}, where binary word encoding is used to incorporate relationship between pixels and their neighbours. This enables LBP to detect patterns in the image irrespective of contrast variations. %The proposed methods speed up the process by utilizing either pixel-wise implementation or filter-based approach\harish{Need to explain difference between pixel-wise and filter-based}. Both of these approaches achieve similar results running at varying passes. %LBP works by extracting features from grayscale image. It thresholds a certain region of pixels using the central pixel’s grayscale value and defines the result as binary patterns. These patterns are further used to label pixels in the image. For analytical purposes, a histogram of these pixels is then formed which represents the feature descriptor.. 
LBP feature extractor is known for its efficiency in utilizing the computation power, but its effectiveness reduces with an increase in noise in the image \cite{ojala2002multiresolution}. Another commonly used method to extract radiomics features is Histogram of Oriented Gradient (HOG) \cite{song2019noninvasive} \cite{zhang2018predictive} %\cite{abbasi2017detection}. %HOG feature descriptor is commonly used for detection of objects in images. 
where the number of oriented gradient occurrences in certain image regions are counted to create a histogram. %Object shapes and edges are critical components in regard to working of HOG. 
Depending on the application, different regions can be used  to capture local shape and edge information from the images, which is further converted into a feature vector using the HOG descriptor. It was found that operating on a larger neighbourhood is better when using HOG for MR images due to the low intensity variance~\cite{nanni2016combining}.

\begin{table}[!t]
\caption{Radiomics in neuro-oncology using handcrafted features.}
\scalebox{0.62}{
\begin{tabular}{c|c|c|c}
\hline
\textbf{Method (year)} & \textbf{Features} & \textbf{Classifier} & \textbf{Accuracy/Specificity/Sensitivity)}\\
\hline \hline \\
%IDSS-based Two stage classification of brain tumor using SVM
\cite{polepaka2019idss} (2019) & LBP & SVM & 97.02/94.28/98.48 \\
%\hline
%A hybrid feature extraction approach for brain MRI classification based on Bag-of-words
\cite{ayadi2019hybrid}	(2019)	& Discrete Wavelet Transform/Bag of words & SVM & 100/-/-\\
%\hline
%Glioma detection on brain MRIs using texture and morphological features with ensemble learning
\cite{gupta2019glioma}	(2019)	& Fusion of LBP, GLCM and GLRL	& Ensemble 	& 97.14/-/- \\
%\hline
%A Noninvasive System for the Automatic Detection of Gliomas Based on Hybrid Features and PSO-KSVM
\cite{song2019noninvasive} (2019)	& \thead{GLCM+PHOG+Intensity-Based + \\ Modified CLBP}	& PSO-SVM & 98.36/97.83/99.17 \\
%\hline
%Differentiation of glioblastoma from solitary brain metastases using radiomic machine-learning classifiers
\cite{qian2019differentiation}	(2019)	& \thead{GLCM +GLRL + gray level size zone matrix \\ + first order statistics + shape descriptors}	& SVM + LASSO		& 90/-/- \\

%\hline
%A predictive model for distinguishing radiation necrosis from tumour progression after gamma knife radiosurgery based on radiomic features from MR images 
\cite{zhang2018predictive}	(2018)	& \thead{GLCM + GLRL +HOG + neighbourhood \\
grey-tone difference matrix}	& RUSBoost ensemble classifier		&73.2/-/-  \\

%\hline
%Local spatio-temporal encoding of raw perfusion MRI for the prediction of final lesion in stroke 
\cite{giacalone2018local} (2018)	& LBP	& SVM	& 95/94/96 \\
%\hline
%Diagnosis of Distant Metastasis of Lung Cancer: Based on Clinical and Radiomic Features 
\cite{zhou2018diagnosis} (2018)	& \thead{GLCM + GLRLM features + \\ Gabor descriptor}	& SVM	&71.02/-/- \\
%\hline
%Radiomics analysis allows for precise prediction of epilepsy in patients with low-grade gliomas 
\cite{liu2018radiomics} (2018)	& Multiple hand crafted & Radiomics Nomogram + ROC	& 81.52/-/- \\

%\hline
%Radiomics signature of computed tomography imaging for prediction of survival and chemotherapeutic benefits in gastric cancer	
\cite{jiang2018radiomics} (2018)	& \thead{radiomics signature \\(Lasso-Cox regression)}	& Thresholding	& 95/-/- \\
%\hline
%Combined FET PET/MRI radiomics differentiates radiation injury from recurrent brain metastasis 
\cite{lohmann2018combined}	(2018)	& \thead{statistical+histogram features + \\ GLCM+GLRLM+GLZLM}	& Logistic Regression		&89/96/85 \\
%\hline
%Building CT Radiomics Based Nomogram for Preoperative Esophageal Cancer Patients Lymph Node Metastasis Prediction
\cite{shen2018building} (2018)	& \thead{GLCM + GLRL + Fractal Dimensions \\  + wavelet filtered GLCM}	& Logistic Regression	& 95/-/- \\

%\hline
%Development and Validation of an MRI Based Radiomics Signature for the Preoperative Prediction of Lymph Node Metastasis in Bladder Cancer 
%\cite{wu2018development}	& 2018	& %first order statistics features + shape based features +  statistics based textural features + wavelet features least absolute shrinkage and selection operator (LASSO) for feature selection
%& Logistic Regression		& 89 - - \\

%\hline
%A radiomic signature as a non-invasive predictor of progression free survival in patients with lower grade gliomas	
\cite{liu2018radiomic} (2018)	& \thead{statistical + shape-based \\ + texture + wavelet}  & LASSO Cox regression model	&	82.3/-/-  \\

%\hline
%Classification of normal and abnormal brain MRI slices using Gabor texture and support vector machines	
\cite{gilanie2018classification} (2017)	& Gabor texture descriptor	& SVM	&97.5-92-99\\

%\hline
%Detection of Brain Tumor in 3D MRI Images using Local Binary Patterns and Histogram Orientation Gradient 
\cite{abbasi2017detection}	(2017)	& LBP + HOG	&Random Forest	& 83.77/-/- \\

\hline
%A novel approach to CAD system for the detection of lung nodules in CT images 
%\cite{javaid2016novel} 2016	&Thresholding, morphological operation, K-means clustering	&SVM	&91.65- -  \\

%\hline
%%Brain tumors detection and segmentation in MR images: Gabor wavelet vs. statistical features	
%\cite{nabizadeh2015brain}&2015	&Gabor wavelet 	&SVM	&95.3-93.9-95.4 \\

%\hline
%Brain tumors detection and segmentation in MR images: Gabor wavelet vs. statistical features	
%\cite{nabizadeh2015brain}&2015	&Statistical Features	 &SVM	&96.7-95.7-96.2\\
%\hline

\end{tabular}\label{HC}}
\end{table}
The first use of gray level co-occurrence matrix (GLCM), a statistical method used for texture analysis by examining spatial relationship between pixels, was recorded in 1973 when Haralick \cite{haralick1973textural} used it to generate state-of-the-art results in image classification. It works by counting the number of times a certain pair of pixels in a specific spatial relationship and with similar gray scale values occur in an image. Recently, GLCM has been widely used for extracting features for disease classification \cite{gupta2019glioma} \cite{liu2018radiomic}   \cite{liu2018radiomics} \cite{lohmann2018combined} \cite{shen2018building}   \cite{song2019noninvasive} \cite{wu2018development} \cite{zhang2018predictive} \cite{bagci2013predicting} \cite{buty2016characterization}. %\harish{\sout{GLCM performs statistical analysis on these measures to form a gray-level co-occurrence matrix.}} %GLCM was improved over the years and 14 texture-based features have been proposed using statistical analyses. GLCM algorithm comprising of these features is used in variety of computational and classification applications involving texture of image. Most of the grayscale images have pixels with similar values in specific regions, making them highly corelated. GLCM exploits this behaviour to produce cooccurrence matrix, which in such cases is diagonally distributed. GLCM works by counting the occurrences of a pixel with gray level value i that is in a spatial relationship with a pixel j within a certain region
Another commonly used method, Gray Level Run Length Matrix (GLRL) \cite{singh2016comparison}, works on the principle of connectivity and extracts quantitative information (lengths) of connected pixels in a specific direction. %In an image of size MxN with maximum gray scale value V, GLRL produces an VxN gray level run length matrix.
GLRL has also been widely used for feature extraction in radiomics studies \cite{zhou2018diagnosis}.

%Gabor filter is best suited to extract textural patterns in optical objects. In this study, 
As a special class of frequency and structure based approaches, Gabor filter has shown to be popular texture analysis approach, and it has also been employed to examine MR scans to filter out texture-based features such as, smoothness, kurtosis, entropy, contrast, mean, and homogeneity. %MR images have been convolved over using 2D Gabor filters to extract these features. 
Gabor filter works especially well for images with uniform patterns. Medical images usually possess pixels with low variance of intensity levels and uniform orientation. Hence, Gabor filter may outperform other texture-based descriptors in case it has the capability to encode narrow bands of occurrences and orientations. Gabor filter is also good at examining structure differentiation that are caused by cancerous cells in MR images making it ideal for medical imaging data. For these reasons, these filters have been used for extracting radiomic features in multiple studies  \cite{gilanie2018classification} \cite{nabizadeh2015brain} \cite{zhou2018diagnosis}. Radiomics is also applied successfully in other diagnostic applications, some recent works are summarized in Table \ref{HC} highlighting the features and classifiers used. 
%\noindent\section*{Classification Methods}
After extracting radiomics features by using various descriptors, a classifier assigns a particular class to the patient image. Most methods (Table \ref{HC}) use support vector machine as a classifier. Other methods include least absolute shrinkage and selection operator (LASSO), random forest and logistic regression. It is important to observe here that there is a wide array of descriptors available and hence requires a lot of handcrafting to chose the most appropriate features. An automated system that can learn features from raw input data could help in providing more generalized results for the increasing number of radiology studies.   % which is a widely known classification algorithm, works by plotting features in n-dimensional space to find optimal hyper plane. This process is performed by training the classifier on various data points. After training the classifier, it is applied on testing data. 

\section{Radiomics using Deep Learning}
%Due to recent advances in both imaging and computers, there is rapid increase in the potential use of artificial intelligence in various radiology based imaging tasks \cite{giger2018machine}. 
Recently, the most widely used machine learning techniques are based on deep learning, where various functions are used to transform the data into a hierarchical representation \cite{schmidhuber2015deep}. DL has gained wide attention in image categorization, image recognition, speech recognition and natural language processing, and medical image analysis \cite{anwar2018medical} \cite{yasaka2018deep}. One major advantage of DL is the fact that features are extracted directly from raw data allowing feature learning \cite{kamilaris2018deep}. DL is also found successful in solving complex problems with limited data, using transfer learning wherein a model trained on one type of data is used to train a different complex task \cite{weiss2016survey}. On the flip side, DL is generally known to be more successful in solving problems where large data sets are available \cite{kamilaris2018deep}, although methods that work for limited data are emerging \cite{wong2018building}.  %There are different types of deep neural networks such as convolutional neural networks (CNNs) and recurrent neural networks (RNNs). CNN are useful for image related tasks whereas RNN are used in speech and language processing tasks  \cite{yasaka2018deep}.The analysis and processing of data is one of the main task for computer. In early phase, as amount of data is small so feature extraction is easy as just useful information is filtered out but when the data size increase the existing methods is not reliable. So to solve this kind of problem, artificial neural network particularly the convolutional neural network has turn out to be the preferred technique \cite{wang2019various}.. In 2012, Krizhevsky et al. proposed architecture of CNN that gained attention due to its high performance in the field of image recognition as compared to conventional methods at ImageNet Large Scale Visual Recognition Competition (ILSVRC)  \cite{krizhevsky2012imagenet}.Subsequently numerous models based on deep learning techniques have been developed for image recognition tasks. Recently, one of the applications of deep learning methods with CNNs for radiological images is gaining wide attention \cite{kahn2017images}\cite{dreyer2017machines}.

%\section{Introduction}
 
%They primarily involve ... 
%The rest of the paper is structured as follows ... 
%\section{}
%\subsection{Machine Learning}
%Here, we can divide this into two sub-sections:
%\subsubsection{Hand-crafted Features}
%Discuss different features being introduced and combined.
%\subsubsection{ML classifiers}
%Discuss the different classifiers experimented on and conclude with summarizing which is the most popular.
%\begin{table}
%\caption{Machine Learning in Radiomics.}\label{tab1}
%\begin{tabular}{|l|l|l|l|l|}
%\hline
%Paper Tile &  Year & Features & Classifier & Accuracy (or other metric)\\
%\hline
%IDSS-based Two stage classification of brain tumor using SVM & 2019 & LBP & SVM & 96.51 

%\end{tabular}
%\end{table}

%\subsection{Deep Learning}
%A short paragraph on the current proposed approach to survival analysis, etc and how they are not yet up-to standard. 

%Table 1 shows the comparison of results obtained from different deep learning architectures.As shown in Fig. \ref{fig:my_label}, 

There are two popular approaches (Figure \ref{fig:my_label}) used in DL - training a network and extracting the features to use with a simple machine learning classifier and training an end-to-end network that incorporates the classification/regression task in it's learning. An example of the former is the work by Nie et. al~\cite{nie2019multi}. The authors proposed a multi-channel artificial intelligence enabled radiomics for predicting patient’s survival time in neuro-oncological applications . First, the proposed technique used three-dimensional convolutional neural networks for extracting high level features from multi-modal MR images. In the second step, those features along with patient’s personal details and medical history were fed to an SVM for predicting the survival time. The proposed method achieved state-of-the-art results with an accuracy of $90.66\%$. %The designed technique is more efficient than previous techniques because it does not incorporate KPS features which consumes more cost and time. 
%It has been observed that deep learning based radiomics in neuro oncology could assist radiologists in devising treatment strategies. 
Chang et. al~\cite{chang2018residual} proposed an end-to-end trained residual convolutional network to diagnose isocitrate dehydrogenase (IDH) mutations in people suffering from grade II-IV gliomas. The diagnosis of IDH mutations could assist radiologists in the treatment of patients suffering from gliomas. The network was trained on multi-institutional clinical MRI data and different techniques like random rotation and zooming was used to reduce data over-fitting. %The training on multi-institutional data made it different from previous researches. 
%Training the network required larger data set which was resolved using data augmentation. 
The proposed network gave an accuracy and area under the curve (AUC) of $82.8\%$ and $0.90$ respectively on training data, $83\%$ and $0.93$ on validation data, and $85.7\%$ and $0.94$ on testing data respectively. This artificial intelligence based radiomics is currently considered as the largest study for the prediction of IDH mutations. In \cite{lao2017deep} proposed deep learning enabled radiomics for survival prediction of patients suffering from glioblastoma multiforme. The proposed technique used transfer learning for predicting patient’s survival. The features were extracted from MR images using conventional and deep learning methods. The features extracted from deep learning were fed to LASSO Cox model for predicting patient’s survival. The proposed technique also required demographic information such as age and Karnofsky Performance Score. The technique has some limitations as it was designed for small dataset and, also the relation between features and patient’s genetic details were not investigated. The results showed that deep learning based radiomics achieved better prognosis than conventional machine learning based radiomics.

There are various methods reported in literature related to brain diseases that are based on both conventional features and DL based methods \cite{altaf2018multi} \cite{hussain2017brain} \cite{hussain2018segmentation} \cite{farooq2017artificial} \cite{ateeq2018ensemble}. In \cite{zhao2018deep}, authors combines fully convolutional neural network with conditional random field (CRF). The technique used image patches for training fully convolutional neural network and 2D image slices: coronal, sagital and axial, for training CRF as recurrent neural network. Then image slices were used to fine tune both networks. The experiments were carried out on BraTS 2013, 2015 and 2016 data sets \cite{menze2014multimodal} \cite{bakas2017advancing}. This study trained three segmentation models using both image patches and slices, and it has been observed that slice by slice segmentation was computationally more effective than segmentation using image patches.  This method worked well for 2D images but did not perform well for 3D volumes. % limitation of this technique is that it worked well for 2D data, but it did not incorporate 3D data.
Cascaded anisotropic convolutional neural networks were employed to segment multi-class brain tumor \cite{wang2017automatic}. The developed technique treated all three classes (core, enhancing, and whole) separately, and three different network architectures were designed and concatenated. %The first network took a 3D volume as input and segmented the whole tumor, which was then fed to second network for segmenting tumor core, and then the third network for segmenting the enhanced region. 
Anisotropic network was designed to resolve model complexity arising from the use of large receptive fields. Residual connections were employed for robust training and segmentation performance. The model was tested on BraTS 2017 dataset \cite{bakas2017advancing} and achieved the dice scores (DSC) of 0.7831, 0.8739, and 0.7748 for enhancing, whole, and core tumor regions respectively. %The validation dataset achieved DSC of 0.9050, 0.8378 and 0.7859 for whole tumor, tumor core and enhanced region respectively. 
The experiments showed that this setup has made training easier and reduced false positives. But this technique is not end-to-end and consumes more time in training and testing than other techniques. 

\begin{table}[!t]
\caption{DL based radiomics approaches in neuro-oncology.}
\scalebox{0.62}{
\begin{tabular}{c|c|c|c|c}

\hline
\thead{\textbf{Method (Year)}} & \textbf{Dataset} & \textbf{Architecture} & \textbf{Task} & \textbf{Performance parameter} \\ \hline \hline %\begin{tabular}[c]{@{}l@{}}Accuracy\\ 			(DSC)\end{tabular}                                                             & Dataset                                                               \\
\thead{Nie et al. (2019)} & Clinical (Glioma) Images  & CNN + SVM   &  Survival prediction & Accuracy: 90.66\%              \\ 
\thead{Chang et al. (2018)} &  Clinical MR Images & Res-Net  & IDH phenotype prediction & \thead{Accuracy: 89.1\% \\(AUC = 0.95)}              \\
\thead{Zhao et al. (2018)} & BraTS  & CNN+ CRF-RNN   & Tumor segmentation & \thead{DSC: Whole-0.82, \\ Core-0.72, Enhanced-0.62}  \\
\thead{Wang et al. (2017)} & BraTS 2017  & Cascaded CNN & Tumor segmentation & \thead{DSC: Whole-0.87, \\ Core-0.77, Enhanced-0.78}  \\
\thead{Alex et al. (2017)} & BraTS 2017 & CNN + Texture Features & Tumor segmentation & \thead{DSC: Whole-0.83, \\Core-0.69, Enhanced-0.72}                 \\
\thead{Havaeiet al. (2017)} & BraTS 2013 & Cascaded CNN & Tumor segmentation & \thead{DSC: Whole-0.81, \\Core-0.72, Enhance-0.58}   \\
\thead{Lao et al. (2017)}& Clinical data & CNN + LASSO Cox & Overall survival & C-index=0.739 \\
\thead{Liu et al. (2017)} & BraTS 2015 & CNN & Tumor segmentation & DSC: Core-0.75, Enhanced-0.81                \\
%Kamnitsas et al. 2016 & CNN with Residual connections &  Whole:0.914 &Core: 0.831 &Enhance:0.794&BraTS 2015,2016          \\
\thead{Kamnitsas et al. (2017)} & BraTS 2015 & 3D-CNN + CRF & Tumor segmentation& \thead{DSC: Whole-0.75, \\Core-0.72, Enhanced-0.898}  \\ \hline
%Pereira et al. 2016 & CNN with smaller kernels &  Whole:0.88, 0.78 &Core:0.83, 0.65 &Enhance:0.77. 0.75& BraTS 2013, 2015         \\
%Dvorak et al. 2015 & Patch based CNN &  Whole:0.83 &Core: 0.75 &Enhance: 0.77& BraTS 2015 \\ \hline              
\end{tabular}
\label{DL}
}
\end{table}

A 23-layered fully convolutional neural network was proposed for segmentation of gliomas from MRI \cite{alex2017automatic}. %Fully convolutional neural network was considered as the best option for semantic segmentation, because it classifies the pixels in a single forward pass. 
Texture analysis including first order texture features and shape-based features was used for the prediction of patient’s survival. The designed algorithm was trained on 2D slices extracted from patient’s MRI volume. The proposed network gave the survival prediction accuracy of $52\%$ and $47\%$ on BraTS 2017 validation and testing dataset respectively. The achieved DSC on BraTS 2017 for whole tumor, tumor core and enhanced region was 0.83, 0.69 and 0.72, respectively. %It has been observed that devised technique used single convolutional network and achieved remarkable results.
A novel CNN architecture was proposed which incorporated both dual pathway and cascaded architecture for radiomics in neuro-oncology \cite{havaei2017brain}. The output of cascaded architecture was fed to dual pathway network improved prediction accuracy. The convolutional neural network predicts labels independent of its neighboring pixels which limits its capability for producing accurate results. The cascaded architecture output made it possible for the proposed CNN to incorporate the influence of neighboring pixels. This variation of convolutional neural network increased the speed by 40 folds and incorporated both local and global features. The fully connected layer of the proposed network architecture was designed in convolutional manner. Two phase training technique was used for accurate delineation of brain tumor and it was tested on BraTS 2013 dataset. The proposed architecture worked well for two-dimensional data but slows down in case of three-dimensional data. 
 
An algorithm was devised using convolutional neural network for segmentation of brain metastases from MRI \cite{liu2017deep}. Image patches were fed to the network for voxel-wise classification which made the setup efficient for segmenting small lesions. %The proposed architecture worked for both mono-modality and multi-modality MR imaging. 
Although the network was designed for mono-modality imaging, nonetheless it was also tested on multi-modality dataset (BraTS),  %The results were compared with deep-medic for both mono-modality and multi-modality data, and it yielded better dice scores. 
where DSC values of 0.75 and 0.81 were achieved on core and enhanced tumors, respectively. The network was trained on pre-defined parameters which made it more robust. The performance of this network architecture could be improved by readjusting patch size and hyper-parameters, however. This AI-enabled radiomics in neuro-oncology could help in treatment strategy planning for brain metastases.
In \cite{kamnitsas2017efficient}, authors proposed \textit{deepmedic} platform, a dual pathway network incorporating local and global features, for segmenting brain tumors. Conditional random fields were used as a post-processing step to reduce the number of false positives.  
An improvement to deepmedic was proposed using residual connections, and performance was evaluated on a small dataset (BraTS 2015) to make this approach more flexible~\cite{kamnitsas2016deepmedic}. This simplified approach achieved good results on BraTS 2016 as well, where DSC score by using 75\% of the data was 91.4, 83.1 and 79.4 for whole tumor, core, and enhanced tumor regions. Table \ref{DL} gives a summary of the methods in segmentation and prediction. Although the results of DL are promising, the methodology suffers with the black box paradigm. The feature learning process is still not transparent and the aim of achieving generalization is still to be achieved. 

\section{Discussion and Conclusion}
%\harish{\sout{The literature shows that}}RaR
AI-enabled radiomics is making significant progress in neuro-oncology and similar applications, with performance better than conventional approaches. It aids radiologists in making an accurate prognosis leading to better treatment strategy. An important consideration is finding the right hand-crafted features, as the results have shown that these features can significantly effect the overall outcome of the method. A possible solution to this impediment is to use DL which is known to learn the right features in an automated fashion, when a reasonable amount of training data is present. It is observed that DL based methods are able to produce state-of-the-art results. Both radiomics and DL fields are currently developing at a very fast pace. It is believed that they will work together in future resulting in the development of AI enabled radiomics that will transform not only prognosis and diagnosis, but also how treatment planning and analysis of disease recurrence works in oncology.

Various tumor types may appear similar on radiology images, but the molecular characteristics of different malignant parts vary. Moreover tumor phenotype changes with the passage of time, hence biopsies cannot provide much information. Hence, personalized medicine predicts different results and more effective treatments, in the light of improved serum, tissue, and imaging bio-markers \cite{subramanyam2016translational}. Radiomics can assist by evaluating the imaging bio-markers that would identify the tumor signature clearly and hence show the tumor function and evolution. These statistics will help multi-disciplinary oncology members to develop a highly personalized curative plan for individuals based on the information of exactly how that specific patient’s cancer is expected to behave. 
%\harish{\sout{A major impact is to be seen in personalized medicine in the near future. }}
Interpretable DL will help in identifying the right radiomic features improving upon the hand-crafted features based methods. For precision and accuracy in this challenging area, more interpretation and explainability is required for the underlying DL-based models. 
\bibliographystyle{splncs04}
%\bibliography{thebibliography}

%\bibliographystyle{plain}
\bibliography{ref}


%Author, F.: Article title. Journal \textbf{2}(5), 99--110 (2016)

%Author, F., Author, S.: Title of a proceedings paper. In: Editor,
%F., Editor, S. (eds.) CONFERENCE 2016, LNCS, vol. 9999, pp. 1--13.
%Springer, Heidelberg (2016). \doi{10.10007/1234567890}

%Author, F., Author, S., Author, T.: Book title. 2nd edn. Publisher,

%Author, A.-B.: Contribution title. In: 9th International Proceedings
%on Proceedings, pp. 1--2. Publisher, Location (2010)

%\end{thebibliography}
\end{document}